\DeclareUnicodeCharacter{0303}{\~}
\documentclass[a4paper,11pt]{article}
\usepackage{pos}

\title{Prospects for Galactic transient sources detection with the Cherenkov Telescope Array}
 \ShortTitle{Galactic transients with CTA}

\author*[a,b]{Alicia L{\'o}pez-Oramas}
\author[c]{A. Bulgarelli}
\author[d]{S. Chaty}
\author[e]{M. Chernyakova}
\author[f]{R. Gnatyk}
\author[f]{B. Hnatyk}
\author[g]{D. Kantzas}
\author[g]{S. Markoff}
\author[e]{S. McKeague}
\author[h]{S. Mereghetti}
\author[i]{E. Mestre}
\author[c]{A. di Piano}
\author[j]{P. Romano}
\author[k]{I. Sadeh}
\author[f]{O. Sergijenko}
\author[h]{L. Sidoli}
\author[l]{A. Spolon}
\author[k]{E. de O{\~n}a Wilhelmi}
\author[m]{G. Piano }
\author[n]{L. Zampieri}
\forColl{CTA} % W/O "Collaboration"

\affiliation[a]{Inst. de Astrof\'isica de Canarias, La Laguna, Spain}
\affiliation[b]{Universidad de La Laguna, Dpto. Astrof\'isica, La Laguna, Tenerife, Spain}
\affiliation[c]{INAF-OAS Bologna, Italy}
\affiliation[d]{University of Paris and CEA Paris-Saclay, France }
\affiliation[e]{School of Physical Sciences and CfAR, Dublin City University, Ireland}
\affiliation[f]{Taras Shevchenko National University of Kyiv, Ukraine}
\affiliation[g]{API/GRAPPA, University of Amsterdam, the Netherlands}
\affiliation[h]{INAF-IASF Milano, Italy }
\affiliation[i]{ICE-CSIC, Barcelona, Spain}
\affiliation[j]{INAF-Osservatorio Astronomico di Brera, Milano, Italy}
\affiliation[k]{DESY-Zeuthen, Zeuthen, Germany}
\affiliation[l]{Universit\`a di Padova and INFN, Padova, Italy}
\affiliation[m]{INAF-IAPS Roma, Italy}
\affiliation[n]{INAF-Astronomical Observatory of Padova, Italy}

\emailAdd{alicia.lopez@iac.es}
\abstract{Several types of Galactic sources, like magnetars, microquasars, novae or pulsar wind nebulae flares, display transient emission in the X-ray band. Some of these sources have also shown emission at MeV--GeV energies. However, none of these Galactic transients have ever been detected in the very-high-energy (VHE; E$>$100 GeV) regime by any Imaging Air Cherenkov Telescope (IACT). The Galactic Transient task force is a part of the Transient Working group of the Cherenkov Telescope Array (CTA) Consortium. The task force investigates the prospects of detecting the VHE counterpart of such sources, as well as their study following Target of Opportunity (ToO) observations. In this contribution, we will show some of the results of exploring the capabilities of CTA to detect and observe Galactic transients; we assume different array configurations and observing strategies.}

\FullConference{37$^{\rm{th}}$ International Cosmic Ray Conference (ICRC 2021)\\
		July 12th -- 23rd, 2021\\
		Online -- Berlin, Germany}

\begin{document}
\maketitle

\section{Introduction}

Many different types of sources in the Galaxy exhibit transient signals. Such emission can occur due to accretion/ejection processes, such as jets interacting with the interstellar medium (ISM), strong winds and/or outflows. In these scenarios, particles may be accelerated up to relativistic energies, leading to the production of high-energy (HE, E$>$100~MeV) and likely very-high-energy (VHE, E$>$100~GeV) radiation, via leptonic or hadronic processes. While many Galactic sources show variable periodic signals, we are interested in those that display irregular unpredictable emission at different wavelengths. 
%Some of these sources are microquasars. These are compact objects, either neutron stars (NS) or black holes (BH), where accretion from a companion star can lead to the formation of a jet. Another example is that of magnetars. The latter are highly magnetized NSs that can display outbursts and flares at different timescales. Finally, we consider pulsar wind nebulae (PWNe). These are sources that exhibit relativistic outflows, which are driven by the energy loss of a rotating NS. 

Several types of Galactic transients exhibiting HE radiation have been detected in the past years by satellites such as \textit{Fermi}-LAT and AGILE. In 2011, AGILE discovered enhanced MeV emission from the Crab Nebula pulsar wind nebula (PWN) \cite{Tavani2011Sci...331..736T}, confirmed by  \textit{Fermi}-LAT\cite{Abdo2011}. This indicates that this source, considered the VHE standard candle, is actually variable at lower energies. Recently, a Galactic magnetar (SGR~${\text{1935\;+2157}}$) has been associated with a Fast Radio Burst (FRB) for the first time \cite{Andersen2020}. This event was not detected in the gamma-ray regime. However, the \textit{Fermi}-LAT has recently discovered GeV emission from an extragalactic magnetar, located in the Sculptor galaxy, which occurred during a giant flare \cite{Fermi2021}. Microquasars, which are binary systems hosting compact objects -either neutron stars (NS) or black holes (BH)- accreting from a companion star, can lead to the formation of a jet. Some microquasars have also been detected in the HE regime \cite{Abdo2009,AGILE2010ApJ...712L..10S,Zanin2016}. Additionally, novae, which are explosions associated to a white dwarf in a close binary system, have been detected in the MeV regime \cite{FermiNova}. Finally, transitional millisecond pulsars (tMSPs) are pulsars in a binary system, which change from an accretion to a radio loud phase; such tMSPs have also been detected in the MeV energy range \cite{ray2012radio}. 
% such objects are also potentially interesting targets for VHE observatories.

%Other sources, like Supergiant Fast X-ray Transients, even if highly variable in the X-ray domain, have never been detected in the HE range.

The current generation of Imaging Air Cherenkov Telescopes (IACTs) are H.E.S.S.~\cite{Aharonian:2006pe}, MAGIC~\cite{2016APh....72...76A}, and VERITAS~\cite{Holder:2006px}. They have successfully discovered more than 200~VHE gamma-ray sources, both of galactic and extragalactic origin. Highlights in the case of Galactic sources include the discovery of gamma-ray binaries with variable emission \cite{Aharonian2005, AharonianLS5039, Albert2006, Aharonian2007, Abeysekara2018p}, pulsations in  the GeV--TeV domain in the emission of pulsars, such as the Crab \cite{MAGICCrabPulsar}, Vela \cite{VelaPulsar} or Geminga \cite{VelaPulsar} or emission from the Galactic centre, revealing it as the first PeVatron (source of cosmic rays with PeV energies) in the Galaxy \cite{HESSGalCenter}, among others. IACTs have been aiming at detecting other types of transient emission, such as the aformentioned sources, without success \cite{MAGICmagnetars,LopezOramas2018,MAGICnovae,VERITASnova}.

\begin{figure}[t]
    \begin{center}
        \includegraphics[width=0.7\textwidth]{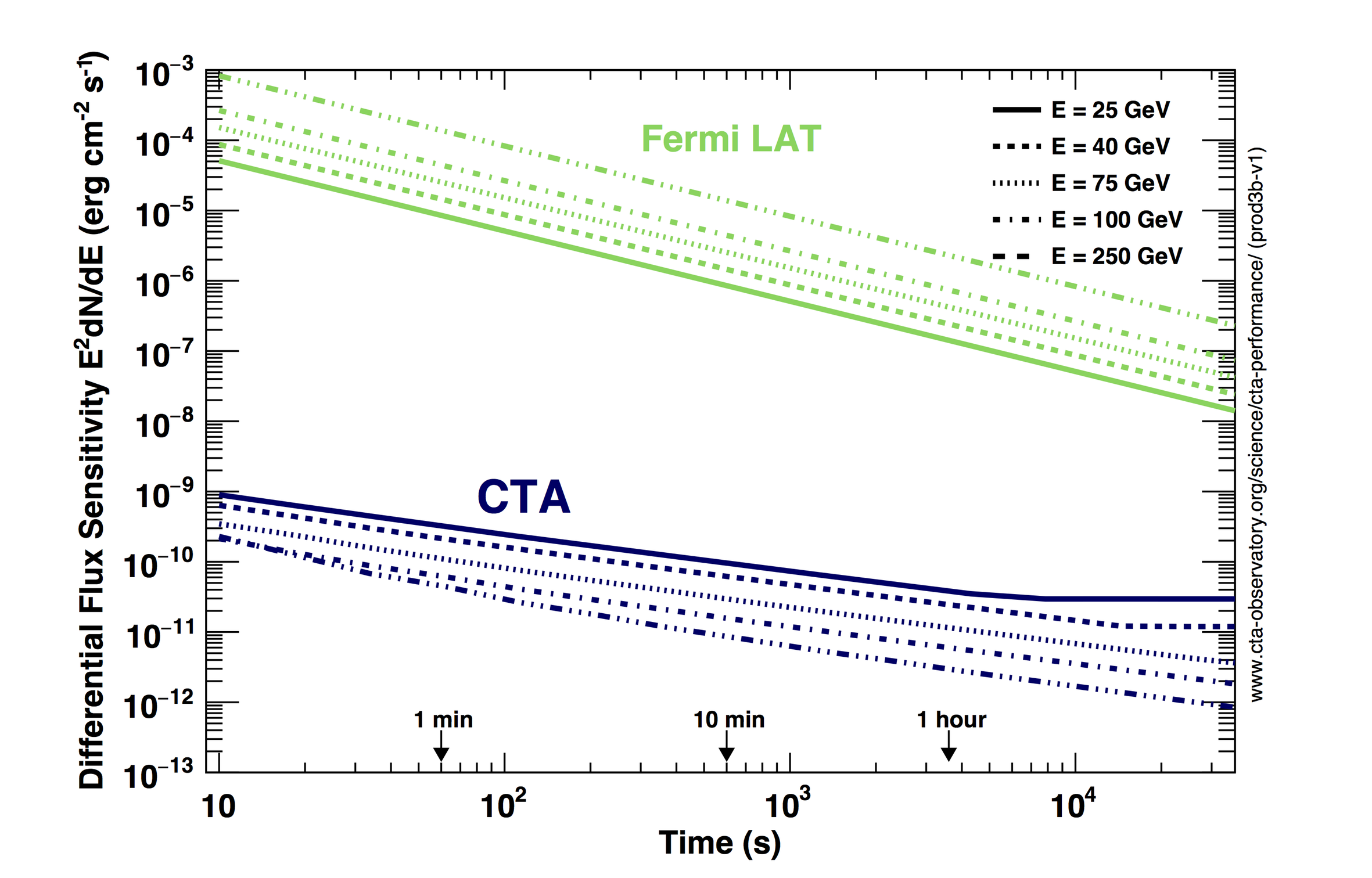}
        \caption{The differential flux sensitivities of CTA and of \emph{Fermi}-LAT for different energies as a function of observation time.}
        \label{fig:CTAsensitivity}
    \end{center}
\end{figure}

The Cherenkov Telescope Array (CTA) will be the next-generation ground-based gamma-ray observatory \cite{Acharya2019}. It will be the premier facility for VHE, multi-messenger, and transients astrophysics in the next decade. CTA will comprise two observatories, one in the Northern (Observatorio Roque de los Muchachos, La Palma) and one in the Southern hemisphere (Paranal, Chile). CTA will be able to perform unprecedented observations of VHE transient sources, covering an energy range from 20~GeV to more than 100~TeV. The larger effective area of CTA, compared to the current generation of IACTs, will result in high sensitivity\footnote{CTA sensitivity: https://www.cta-observatory.org/science/cta-performance/} at short timescales (see Fig.\ref{fig:CTAsensitivity}) \cite{Fioretti2019}. This will enable unique studies of the multi-messenger and transient sky \cite{2021arXiv210603621B}, which is one of Key Science Projects of the observatory~\cite{Acharya2019}.
It is correspondingly necessary to carefully understand the capabilities of CTA to detect such sources (see Carosi, ICRC2021, id.833). In many cases, CTA observations will be based on external triggers from various monitoring instruments (X-ray or HE satellites). Serendipitous discoveries can also take place, i.e., as part of the nominal observation of the Galactic plane survey (GPS). The \textit{Science Alert Generation}, which is a real-time very-short timescale (from 1 to 100 seconds) analysis will play a key role in the follow-up of external triggers and in the serendipitous discovery of transient events (see di Piano, ICRC2021, id.156).

While certain objects show persistent emission and/or periodically emit variable radiation, in this contribution we will mainly focus on those emitters on those emitters which display irregular and  unpredictable transient emission at different wavelengths. 
We will summarize the work of the Galactic transients task force of the CTA Consortium, which is focused in understanding the capabilities of CTA for detecting transient sources of Galactic origin. The results shown in this contribution will be discussed in more detail in an upcoming CTA Consortium paper.
%The results shown here are part of an upcoming comprehensive CTA Consortium paper.

\section{Sensitivity in the Galactic plane}

\begin{figure}[th!]
    \begin{center}
        \includegraphics[trim=45mm 60mm 78mm 25mm,clip,width=0.9\textwidth]{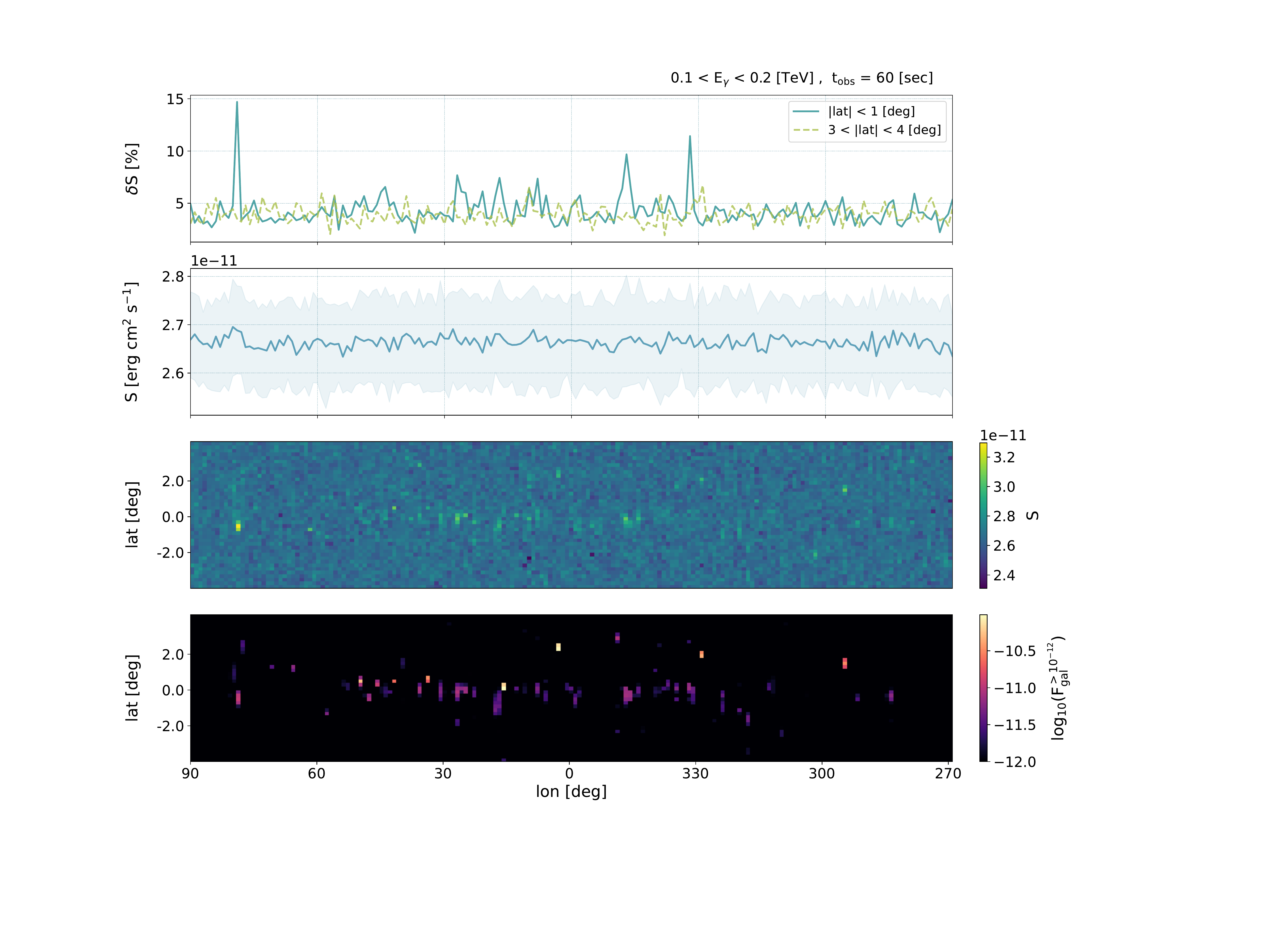}
        \caption{Flux sensitivity ($S$) of CTA-S  within 100--200~GeV for 60~s observation intervals, considering different perspective source locations along the Galactic plane. 
        %We define $S$ as the minimal flux of a source, multiplied by the energy squared, such that the source is detectable at $5\sigma$ significance. 
        %The bottom panel shows a simulation of $F_{\text{gal}}^{>10^{-12}}$, 
        %the Galactic emission above a threshold of ${10^{-12}~\text{erg}\;\text{cm}^2\;\text{s}^{-1}}$, which is derived
        %for different Galactic longitudes, \textit{lon}, and latitudes, \textit{lat}.
        The bottom panel shows a simulation of $F_{\text{gal}}^{>10^{-12}}$ , the Galactic emission above a threshold of ${10^{-12}~\text{erg}\;\text{cm}^2\;\text{s}^{-1}}$. The emission is integrated within 0.25$^{\circ}$ radial regions around each position, corresponding to different Galactic longitudes, \textit{lon}, and latitudes, \textit{lat}.
        The next panel above shows the corresponding CTA sensitivity. In the third panel we present the median of $S$ for different longitudes within the range, $-4 < \text{lat} < 4$~deg, where the shaded uncertainty region represents the $1\sigma$ variance of $S$. Finally,  the top panel shows the relative $1\sigma$ variance, $\delta{S}$, derived for two ranges in latitude, as indicated. 
        The variance away from the Galactic Plane ($3 < |\text{lat}| < 4$~deg)
        represents the intrinsic statistical uncertainty of the sensitivity calculation.
        The variance in the inner Galactic region ($|\text{lat}| < 1$~deg) includes the intrinsic uncertainty, as well as the additional effect of the steady Galactic foregrounds, which are concentrated in this region. }
        \label{fig:CTAsensitivityMap}
    \end{center}
\end{figure}

It is important to understand the performance of the CTA Northern and Southern arrays, especially in the Galactic plane, where many TeV sources are located.
The sensitivity of the Southern array is illustrated in Fig.~\ref{fig:CTAsensitivityMap} for different perspective source locations, and is defined as the minimal flux of a source, such that the source is detectable at $5\sigma$ significance within a given energy range. We define the variable \textit{S}, which stands for the sensitivity multiplied by the energy squared.
%The sensitivity, \textit{S}, is defined as the minimal flux of a source, multiplied by the energy squared, such that the source is detectable at $5\sigma$ significance within a given energy range. We have simulated the sensitivity in the Galactic plane using \textit{ctools}\footnote{http://cta.irap.omp.eu/ctools/index.html} \cite{ctools}, a software created for the analysis of scientific data of CTA. 
For the current example, we estimate the performance of CTA for short observation intervals of 60~seconds, within the 100--200~GeV energy range.
%utilize simulations of the Galactic plane ...... (\textcolor{red}{cite}), using \textit{ctools}\footnote{http://cta.irap.omp.eu/ctools/index.html}, a software created for the analysis of scientific data of CTA \textcolor{red}{(ctools description and reference.)}
%For the current example, we estimate the performance of CTA for short observation intervals of 60~sec, within the energy range, 100--200~GeV.
% 

As it can be inferred from the figure, upward fluctuations of the sensitivity (worse performance of CTA), are correlated with the simulated Galactic emission.
The flux of the expected steady Galactic foreground in the chosen energy range is mostly
below the level of a few ${10^{-11}~\text{erg}\;\text{cm}^2\;\text{s}^{-1}}$.
This is of the same order as the nominal sensitivity of the observatory in the absence of foregrounds. Correspondingly, the overall degradation in sensitivity for detection of new sources is not significant; at worst, it amounts to a relative increase of the flux threshold of $5\text{--}10\%$, and only when coinciding with strong Galactic emitters.

As part of our upcoming publication, we will explore the performance along the Galactic plane for different energy ranges and observation times. Given the assumed properties of known Galactic sources, the performance is not expected to significantly diverge from the nominal capabilities of CTA, which are shown in
Fig.~\ref{fig:CTAsensitivity}.

% 
%\begin{figure}[t]
%    \begin{center}
%        \includegraphics[width=\textwidth]{CTA-Short-TimeScaleSensitivity-201804504-10h-1}
%        \caption{ The differential flux sensitivities of CTA and of \emph{Fermi}-LAT for different energies as a function of observation time \cite{CTASensitivity}. }
%        \label{fig:CTAsensitivity}
%    \end{center}
%\end{figure}

\section{Detecting Galactic transients with CTA}

We have tested the capabilities of CTA to detect transient VHE emission for different kinds of Galactic sources. We have assumed different array configurations, namely the full CTA Northern (CTA-N), and CTA Southern (CTA-S) arrays, as well as different sub-arrays of CTA telescopes, different observing strategies, etc. Below we illustrate our findings with three source examples, microquasars, flaring PWNe and tMSPs, which are sources that are known to show MeV emission.

\subsection{Microquasars}

Microquasars are binary systems composed of a compact object (BH or NS) which accretes matter from a companion star. Depending on the mass of the companion, they can be divided into high-mass or low-mass systems. Microquasars normally present an accretion disk, and can produce collimated jets of plasma. These jets are normally active during the so-called \textit{hard state}. 

We have studied three microquasars located in the Cygnus region: the two high-mass systems Cyg X-1 and Cyg X-3 and low-mass binary V404 Cyg. Both Cyg X-1 and Cyg X-3 are sources of HE gamma rays \cite{Abdo2009,AGILE2010ApJ...712L..10S, Zanin2016}, although the nature of their emission mechanisms (hadronic or leptonic) is still uncertain. Correspondingly, the emission may be attributed to a jet interacting with the surrounding interstellar material (ISM), or from the coronal region of the accretion flow. The latter scenario is quite interesting, indicating that microquasars are potential accelerators of cosmic rays via magnetic reconnection \cite{Khiali2015}. 

Searches for both transient and persistent VHE emission from microquasars have not resulted in detections, either from the binaries themselves, or from jet--ISM interactions \cite{Aleksic2010,Ahnen2017}. Only, a hint of transient emission from Cyg X-1 was observed by the MAGIC telescopes in 2006, over an 80-minute observation interval \cite{Albert2007}. The low-mass binary V404 Cyg displayed a major flaring episode in X-rays in June 2015, after 26 years in quiescence; unfortunately, only 4$\sigma$ evidence was observed by \textit{Fermi}-LAT \cite{Loh2016}, falling short of a correlated detection. No VHE emission was observed either \cite{MAGICV404}.
% 
%Some models predict TeV emission under efficient particle acceleration within jets, as well as a strong hadronic jet component in the case of low-mass binaries. However, no detection was seen by MAGIC at VHE for this source \cite{MAGICV404}. %\textcolor{red}{Sadeh: I hope i didn't change the meaning of the end of this paragraph...}

We have tested different scenarios in which we would expect transient and persistent emission from any of these binary systems. CTA will not be able to detect a flare from V404 Cyg. However, our simulations indicate that CTA will detect transient emission in both Cyg X-1 and Cyg X-3, where we wish to highlight the expected detection of Cyg X-1 in only 0.5~h, as shown in Fig.\ref{fig:GalacticTransients}.

For a small number of microquasars, the jets are persistent, such as for SS433. This microquasar has been detected in the MeV range by \textit{Fermi}-LAT \cite{Fermi2020}; it is the only one also detected in the TeV range, as reported by HAWC \cite{Abeysekara2018}. The observed extended TeV emission is due to the interaction between the jet and the surrounding nebula (the so-called \textit{lobes}), while the central binary remains undetected. To date, this source has  not been detected by an IACT \cite{Ahnen2018}, which would have helped to fill in the as-yet unexplored GeV--TeV energy range between \textit{Fermi}-LAT and HAWC. Our simulations indicate that both the CTA-N and CTA-S arrays will be able to detect the central binary SS433 and its lobes with high significance. CTA will thus provide crucial insight into the emission models of microquasars, and advance our understanding of jet formation.

\subsection{Flares from PWNe}

The discovery of MeV flares from the Crab Nebula \cite{Tavani2011Sci...331..736T, Abdo2011} revealed that PWNe can also display transient variable emission, even if most of these systems are detected as steady sources. These flares, however, have never been detected in the GeV--TeV regime. The CTA observatory will advance our understanding of the origin of the Crab Nebula flares. The low-energy threshold of CTA will allow sampling of the \textit{Fermi}-LAT spectral shape. Complementary observations in the TeV regime will be used to explore a possible inverse Compton (IC) component of the emission, which might arise via the off-scattering of the MeV flares. 

\begin{figure}[!h]
\centering
\includegraphics[width=\textwidth]{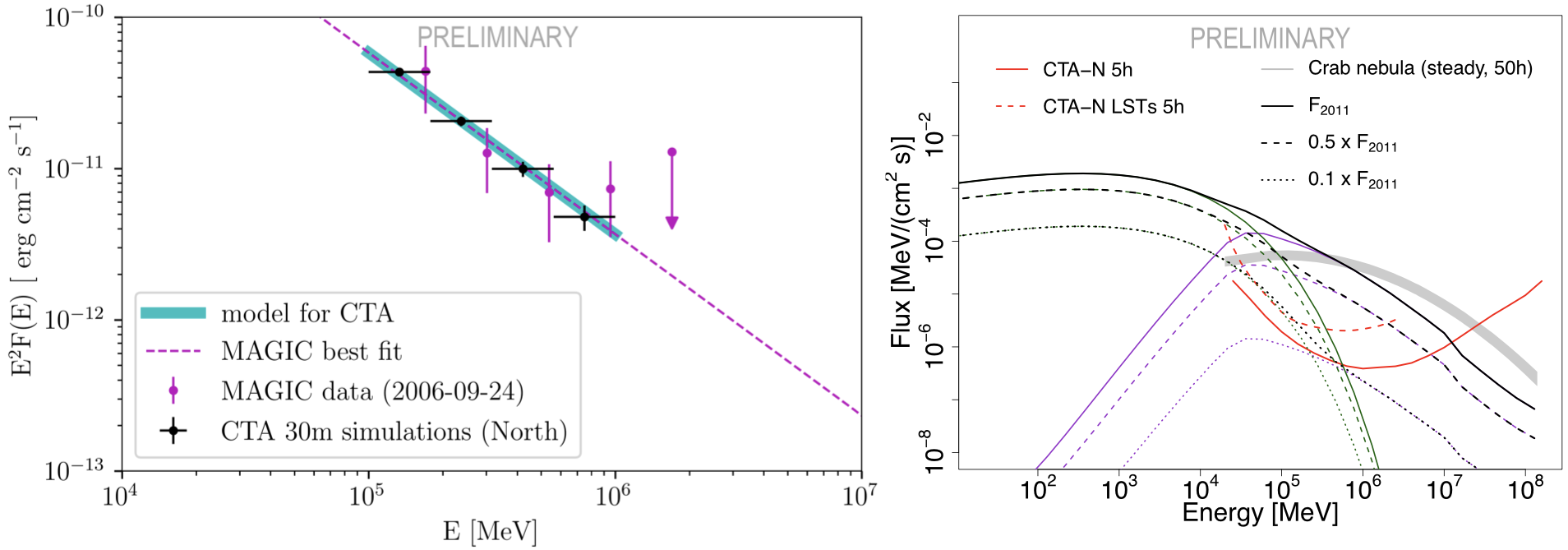}
\caption{\small{\textit{Left}: Spectral energy distribution of Cyg X-1 during a flaring episode similar to that reported in \cite{Albert2007} (magenta points). CTA will detect this binary in 30 minutes of observation (black points; model in cyan). \textit{Right}: simulation of different flaring models (black lines), result of the synchrotron (green) and IC (purple) contributions. The red lines correspond to the sensitivities of CTA-North and the 4 Large Size Telescopes (LSTs) of CTA-North (for 5 h). The steady Crab spectrum is plotted for comparison (gray shaded area). Figures extracted from the CTA Consortium paper on Galactic transients (in prep.)}} \label{fig:GalacticTransients}
\end{figure}

We have tested how the Northern CTA array will detect such flaring episodes, using two array configurations. The first is the full CTA array (CTA-N), including four Large Size Telescopes (LSTs) and 15 Medium Size Telescopes (MSTs). The second configuration is a partial sub-array of telescopes, composed exclusively of four LSTs (CTA-N LSTs), which dominate the low-energy sensitivity range of the observatory. Our conclusion is that CTA will be able to detect these flares in less than 5~h, even if only the reduced 4-LST sub-array is available. The results for detecting flaring episodes with different flux levels is show in Fig.\ref{fig:GalacticTransients}. %As shown, CTA will be able to detect these flares in less than 5 h, even if only observing the 4-LSTs sub-array.

\subsection{Transitional millisecond pulsars (tMSPs)}
tMPS are binaries composed of a low-mass star and a pulsar with a millisecond-duration period. These systems change from an accretion-powered phase to a radio loud phase. Three tMSPs have been detected at HE by \textit{Fermi}-LAT during the accretion-powered phase \cite{ray2012radio}. Transitions between the two states can occur on timescales from days to weeks, producing variability in the whole electromagnetic spectrum. We have tested whether CTA could detect emission from two systems, PSR J1023+0038 and XSS J1227-48538, which are detected with \textit{Fermi}-LAT in the 0.1--10 GeV energy range during the accretion phase. We show that the full CTA array will be able to detect persistent emission from these sources by performing long integration-time observations ($>$50h). 
%No transient emission will be detected from this kind of sources.

\section{Summary}

CTA will perform real-time TeV studies of the variable Galactic sky. Its unique sensitivity to short-timescale events and its low energy threshold make this observatory a powerful and efficient instrument to detect and discover new transient sources. The observational strategy of CTA includes both follow-up of externally triggered events, and serendipitous discoveries, taking place e.g., while conducting a Galactic plane Survey. CTA will be able to detect microquasars such as Cyg X-1, Cyg X-3 and SS433 in the GeV-TeV domain for the first time. It will also be able to probe the flaring episodes of PWNe (namely the Crab Nebula) at VHE. CTA will detect for the first time the VHE component of tMSPs, by performing dedicated long-time observations ($>$50h) during the accretion state.

The unique capabilities of CTA will also likely result in the detection of other large variety of Galactic transients, such as novae, flares from known gamma-ray/X-ray binaries and magnetars. Serendipitous discoveries are also possible. CTA will play a key role in identifying the nature of new transients by performing follow-up observations of wide field-of-view instruments. CTA will help to reveal their most energetic counterpart, and to unveil the acceleration mechanisms at work. 
The real-time analysis \textit{Science Alert Generation}  will play a key role in the follow-up and observation strategies of externally triggered events and also in the serendipitous discovery of transient events. \\

\small
\textbf{Acknowledgements}: This research made use of ctools, a community-developed analysis package for Imaging Air Cherenkov Telescope data. ctools is based on GammaLib, a community-developed toolbox for the scientific analysis of astronomical gamma-ray data.

%\begin{thebibliography}{99}
%\bibitem{...}
\small
\bibliographystyle{apsrev}
\bibliography{cta_icrc}
%\end{thebibliography}

%% Full authors list (ONLY FOR COLLABORATIONS)
%\clearpage
%\section*{Full Authors List: \Coll\ Collaboration}
%
%\noindent \textbf{Note comment afterwards:} Collaborations have the possibility to provide an authors list in xml format which will be used while generating the DOI entries making the full authors list searchable in databases like Inspire HEP. For instructions please go to icrc2021.desy.de/proceedings or contact us under icrc2021proc@desy.de.\\
%
%\scriptsize
%\noindent
%first.author$^1$, 
%second.author$^2$, 
%third.author$^3$ % .... more names
%and 
%last.author$^{n}$ \\
%
%\noindent
%$^1$first.affiliation.
%$^2$second.affiliation. % .... more affiliation
%$^{m}$last.affiliation.

%\documentclass{article}
%\usepackage[utf8]{inputenc}

%\usepackage{natbib}
%\usepackage{graphicx}
%\usepackage{pos}

\parindent 0pt 

%\begin{document}
\textbf{The Cherenkov Telescope Array Consortium July 2021 Authors}

H.~Abdalla\textsuperscript{1}, H.~Abe\textsuperscript{2},
S.~Abe\textsuperscript{2}, A.~Abusleme\textsuperscript{3},
F.~Acero\textsuperscript{4}, A.~Acharyya\textsuperscript{5}, V.~Acín
Portella\textsuperscript{6}, K.~Ackley\textsuperscript{7},
R.~Adam\textsuperscript{8}, C.~Adams\textsuperscript{9}
% ...

1 : Centre for Space Research, North-West University, Potchefstroom, 2520, South Africa

2 : Institute for Cosmic Ray Research, University of Tokyo, 5-1-5, Kashiwa-no-ha, Kashiwa, Chiba 277-8582, Japan

3 : Pontificia Universidad Católica de Chile, Av. Libertador Bernardo O'Higgins 340, Santiago, Chile

4 : AIM, CEA, CNRS, Université Paris-Saclay, Université Paris Diderot, Sorbonne Paris Cité, CEA Paris-Saclay, IRFU/DAp, Bat 709, Orme des Merisiers, 91191 Gif-sur-Yvette, France

5 : Centre for Advanced Instrumentation, Dept. of Physics, Durham University, South Road, Durham DH1 3LE, United Kingdom

6 : Port d'Informació Científica, Edifici D, Carrer de l'Albareda, 08193 Bellaterrra (Cerdanyola del Vallès), Spain

7 : School of Physics and Astronomy, Monash University, Melbourne, Victoria 3800, Australia

8 : Laboratoire Leprince-Ringuet, École Polytechnique (UMR 7638, CNRS/IN2P3, Institut Polytechnique de Paris), 91128 Palaiseau, France

9 : Department of Physics, Columbia University, 538 West 120th Street, New York, NY 10027, USA

% ...

\end{document}